\documentstyle[12pt]{article}
\def\beq{\begin{equation}}
\def\eeq{\end{equation}}
\def\bea{\begin{eqnarray}}
\def\eea{\end{eqnarray}}
\def\bq{\begin{quote}}
\def\eq{\end{quote}}

\def\HPA{{\it Helv.Phys.Acta} }

\def\HPA{{\it Helv.Phys.Acta} }

\def\gappeq{\mathrel{\rlap {\raise.5ex\hbox{$>$}}
{\lower.5ex\hbox{$\sim$}}}}

\def\lappeq{\mathrel{\rlap{\raise.5ex\hbox{$<$}}
{\lower.5ex\hbox{$\sim$}}}}

\def\Toprel#1\over#2{\mathrel{\mathop{#2}\limits^{#1}}}

\begin{document}
\pagestyle{empty}
\begin{flushright}
{CERN-TH/2003-235}
\end{flushright}
\vspace*{5mm}
\begin{center}
{\bf BOUND STATES IN TWO SPATIAL DIMENSIONS IN THE NON-CENTRAL CASE}
\\
\vspace*{0.5cm}
{\bf Andr\'e MARTIN}\\ \vspace{0.1cm}
Theoretical Physics Division, CERN\\
CH - 1211 Geneva 23, Switzerland\\
and\\
Laboratoire de Physique Th\'eorique ENSLAPP\\
F - 74941 Annecy-le-Vieux, France\\
\vspace{0.8cm}
{\bf Tai Tsun WU}\footnote{Work supported in part by the U.S. Department of
Energy under Grant No.
DE-FG02-84-ER40158}\\ \vspace{0.1cm}
Gordon McKay Laboratory, Harvard University\\
Cambridge, MA 02138-2901, U.S.A.\\
and\\
Theoretical Physics Division, CERN\\
CH - 1211 Geneva 23, Switzerland\\

\vspace*{1.5cm}

{\bf ABSTRACT} \\ \end{center}

\vspace*{2mm}

We derive a bound on the total number of negative energy bound states in a
potential in two spatial dimensions by using
an adaptation of the Schwinger method to derive the Birman-Schwinger bound
in three dimensions. Specifically, counting
the number of bound states in a potential $gV$ for $g = 1$ is replaced by
counting the number of $g_i$'s for which zero
energy bound states exist, and then the kernel of the integral equation for
the zero-energy wave functon is symmetrized.
One of the keys of the solution is the replacement of an inhomogeneous
integral equation by a homogeneous integral
equation.
\vspace*{1.3cm}

\begin{flushleft} CERN-TH/2003-235 \\
August 2003
\end{flushleft}
\newpage

\setcounter{page}{1}
\pagestyle{plain}

\section{Introduction}

In a previous paper \cite{aaa}, K. Chadan, N.N. Khuri and ourselves (A.M.
and T.T.W.) obtained a bound on the number of bound
states in a two-dimensional central potential. This bound has the merit
that, for a potential $gV$, the coupling constant
dependence for large $g$ is optimal, i.e., the same as the one of the
semiclassical estimate \cite{bb}. Previous work on the
subject was done by Newton \cite{cc} and Seto \cite{dd}.  We also obtained
a bound for the non-central case,
but only by using a rather brutal method which consists of replacing the
potential by a central potential
which is defined, after choosing a certain origin, by
\bea
V_c(r) & = &  In f \, V(\vec{x}) \nonumber \\
& & |\vec{x}| = r
\label{1}
\eea
Because of the monotonicity of the energy levels with respect to the
potential, putting $V_c$ in our formulae will give a
bound for the potential $V$. However, for potentials with singularities
outside the given origin, this may lead to no bound
at all.  Our attention has been attracted by the fact that in condensed
matter physics problems exist, where
counting the bound states on a surface may be useful \cite{ee}, but where
it is very unlikely that the
potential will be central, even approximately.

In the present paper, we obtain a bound on the number of bound states in a
non-central two-dimensional potential, using an
adaptation of the Schwinger method to derive the Birman-Schwinger bound
\cite{ff} in the three-dimensional case.  The
condition under which we obtain a bound is
\beq
\int \int d^2x \, d^2y |V(x)| (\ell n |x-y|)^2 |V(y)| < \infty.
\label{2}
\eeq
This condition is \underline{non-linear}, just like that of Birman and
Schwinger, but we show in the Appendix that it
follows from the linear conditions
\bea
\int d^2 x & \, \left( \ell n (2+|x|) \right)^2 \,|V(x)| < \infty
\nonumber \\
 {\rm and} \nonumber \\
\int d^2 x & \, V_R (|\vec{x}|) \, \ell n^-|x| < \infty ,
\label{3}
\eea
where $V_R$ is the circular decreasing rearrangement of $|V|$ (for the
definition of $V_R$, see the Appendix).

Condition (2) has already been proposed by Sabatier \cite{ggg}.  Condition
(3) appears in a forthcoming work by N.N. Khuri, A.
Martin, P. Sabatier and T.T. Wu, dealing with the scattering problem. It
has the advantage of showing more clearly what kind
of behaviour the potential is allowed to have at short and large distances.

The strategy of Schwinger consists of counting the number of zero-energy
bound states for a potential $gV$ for $0 < g < 1$
instead of the actual number of negative energy bound states for $g = 1$.
In three dimensions these two numbers are equal.
Indeed, let $0 < g_1, g_2, \dots g_n < 1$ be the coupling constants for
which  we have  zero-energy bound states. Each $g_i$
is the origin of a bound state trajectory in the $E-g$ plane, $E_i(g)$,
with $E_i(g_i) = 0$. These trajectories are
monotonous decreasing:
\beq
\frac{dE_i}{dg} = \int \, V \, \psi^2 d^nx,
\label{4}
\eeq
by the Feynman-Hellmann theorem, but
\beq
g_i \int \, V \, \psi^2 d^n x = E - \int \,|\nabla \psi |^2 \, d^n x < 0 \;
{\rm for} \; E < 0.
\label{5}
\eeq

This shows that the number of negative-energy bound states is exactly the
same as the number of $g_i$'s $< 1$.  At the
crossing of any pair of trajectories there is no problem because of their
monotonicity.

The same result holds in two dimensions with one modification: any
attractive potential (i.e., $\int \, d^2x \, V < 0$),
has a bound state for arbitrarily small $g$, with a binding energy going to
zero for $g \to 0$ like $- \exp - (C/g)$
\cite{aaa}. At $E = 0$ it disappears and is not included in Schwinger's
accounting, so we have to add one unit.

Since we only want a \underline{bound} on the number of bound states, we
can always replace $V(y)$ by $-V^-(y)$:
\bea
V^-(y) & = & 0 \, {\rm for} \, V > 0 \nonumber \\
V^-(y) & = & -V(y) \, {\rm for} \, V \leq 0
\label{6}
\eea
Using $-|V(y)|$ instead of $-V^-(y)$ gives a more crude bound.

It can be shown that the general solution of the zero-energy Schr\"odinger
equation
\beq
- \Delta \psi - V^- \psi = 0
\label{7}
\eeq
in the equivalent integral form
\beq
\psi (x) = C - \frac{1}{2\pi}  \, \int \, \ell n k_0 |x-y| V^-(y) \psi (y) d^2y
\label{8}
\eeq
has a general asymptotic behaviour for $|x| \to \infty$
\beq
\psi (x) - C \sim - \ell n k_0 |x| \; \frac{1}{2\pi}  \; \int \, V^-(y)
\psi (y) d^2 y + o (1)
\label{9}
\eeq
under the condition
\beq
\int \, V^-(y) \, \left( \ell n (2+|y|) \right)^2 \, d^2y < \infty
\label{10}
\eeq
which follows from condition (1) as shown in the Appendix.

Zero-energy bound states are characterized by the fact that $\psi$ is
\underline{bounded}.  Hence we get  the necessary
condition:
\beq
\int \, V^- (y) \psi (y) d^2y = 0
\label{11}
\eeq
Now we have two possibilities:\\
\begin{itemize}
\item
I. At infinity $\psi (\vec{x}) \to 0$ and hence $C = 0$ and those bound
states wave functions satisfy a
\underline{homogeneous} integral equation
\beq
\psi_i (x) = - \frac{g_i}{2\pi} \int \, d^2 y \; \ell n (k_0 |x-y|) \,
V^-(y) \psi_i (y)
\label{12}
\eeq
(notice that the scale factor $k_0$ disappears because of condition (11)).

This is what happens in the case of a central potential for a non-zero
azimuthal angular momentum $m$.

\item
II. At infinity, $\psi (\vec{x}) \to C$, with $C \not= 0$. In this case,
the bound state wave functions satisfy an
inhomogeneous integral equation.  This case has been described in Ref.
\cite{hh}, where it is shown that for a central
potential in two dimensions, the $m = 0$ phase shift has the universal
behaviour
\beq
\delta (k) \sim \frac{\pi}{2 \ell n k } , \; {\rm for} \; k \to 0
\label{13}
\eeq
\underline{except} if there is a zero-energy bound state of type II.  Then
\beq
\delta (k)  \ell n k  \to 0 .
\label{14}
\eeq
In Ref. \cite{hh}, a much stronger result is stated. This much stronger
result, however, holds only for a very rapidly
decreasing potential.
\end{itemize}

\section{Counting Bound States in Case I}

Following Schwinger, we symmetrize the kernel of the integral equation:
\beq
\phi_i (x) = g_i \int \,K (x, y) \phi_i (y) d^2 y
\label{15}
\eeq
with
\bea
\phi_i (x) & = & \sqrt{V^- (x)} \; \psi_i(x) \nonumber \\
K (x , y) & =  &- \frac{1}{2\pi} \; \sqrt{V^- (x)} \; \ell n \, k_0 |x-y|
\; \sqrt{V^- (y)}.
\label{16}
\eea
If $V^-(x)$ vanishes in some regions, it seems impossible to go back from
$\phi_i$ to $\psi_i$.  However, this can be
remedied by defining
\beq
V^-_{\epsilon}(x) = V^- (x) + \epsilon  \exp - \mu |x| .
\label{17}
\eeq
Since the bounds we shall get are continuous in $V$, we can take the limit
$\epsilon \to 0$ at the end.

$K$ can be written as
\beq
K = \Sigma \, \frac{1}{g_i} \, |\phi_i> <\phi_i | + R
\label{18}
\eeq
$R$ is a sum over states which do not satisfy (11).  The $\phi_i$'s in
themselves do not form a complete set.  If we define
$a$ by
\beq
a(x) =
\frac{\sqrt{V^-(x)}}{\sqrt{\int \, V^- (y) d^2y}}
\label{19}
\eeq
we have
\beq
<a | \phi_i > = 0
\label{20}
\eeq
from property (11), and naturally $<a | a> = 1$.

If we define $\hat{Tr}$, a trace restricted to the $\phi_i$'s, we have
$$\hat{Tr} K = \Sigma \, \frac{1}{g_i} > \sum_{g_i \leq 1} \; \frac{1}{g_i}
> N_I,$$
$N_I$ being the number of bound states of type I.  However, this trace
turns out to be divergent because of the logarithmic
singularity of the kernel (the same happened in Schwinger's original
work!), and we follow Schwinger to iterate the integral
equation (15):
$$
\phi_i(x) = g^2_i \int \, K (x , y) K (y, z) \, \phi_i (z) d^2y d^2z
$$
and then
\beq
\hat{Tr} K^2 = \Sigma \, \frac{1}{g_i^2} > N_I
\label{21}
\eeq
Forgetting the ``hat" on the trace still gives a bound because $K^2$ is a
positive operator (contrary to K!), but this bound
depends on the scale parameter $k_0$ entering in the logarithm. Among the
missing states in $\hat{Tr}$ is the state
$|a>$, orthogonal to the $\phi_i$'s, and this one should be removed from
the complete trace.  In this way, we get
$$N_I <
TrK^2 - <a|K^2|a> ,$$
or more explicitly
\bea
& N_I & <  \frac{1}{(2\pi)^2} \, \int \, V^-(x) (\ell n \, k_0 |x-y|)^2
V^-(y) d^2x d^2y \nonumber \\
& - & \frac{1}{(2\pi)^2} \; \frac{1}{\int V^- (x ) d^2x} \; \int \, V^- (x)
(\ell n \, k_0 |x-z|) \;
V^-(z) (\ell n \, k_0 |z-y| ) \
 V^-(y) \, d^2x d^2y d^2z
\label{22}
\eea
It is visible that the second term is negative as we announced.

Rewriting $N_I$ as
$$
\frac{1}{\int V^- (z) d^2 z} \;
\int d^2x d^2y d^2z\, V^-(x) V^-(y) V^-(z) \;
\left[ (\ell n \, k_0 |x-y|)^2 - \ell n \, k_0 |x-z| \ell n \, k_0 |y-z|
\right ],
$$
we see that (22) is manifestly \underline{independent} of the scale factor
$k_0$.

\section{Counting Bound States in Case II}

At first it would seem that Schwinger's technique will not work because, in
Eq. (8), the constant is not zero and therefore
we deal with an \underline{inhomogeneous} integral equation which can be
written, after the same changes of variables as in
Section 2, given by (16) and (19):
\beq
\phi_i = C_i a + g_i K \phi_i ,
\label{23}
\eeq
with, again,
\beq
<a |\phi_i >  =   0 \;
{\rm and} \;
< a | a>   =   1
\label{24}
\eeq
Equation (24) is precisely the key property which will make it possible to
replace (23) by a \underline{homogeneous equation}.

Again, the $\phi_i$'s corresponding to different $g_i$'s are orthogonal because
\bea
<\phi_i | \phi_j >  =  C_j < \phi_i | a> + g_j < \phi_i | C \phi_j>
\nonumber
\\
  =  C_j < \phi_j | a> + g_i < \phi_i | C \phi_j> \nonumber.
\eea
Hence, from (24):
\beq
\left( \frac{1}{g_i} - \frac{1}{g_j} \right ) < \phi_i | \phi_j > = 0.
\label{25}
\eeq

Let us call ${\cal S}$ the Hilbert space associated to the integral
equation (23), and construct a new Hilbert space by
removing the element $a$:
\beq
{\cal S} ={\cal S '} \oplus \{ a \}
\label{26}
\eeq
We want to define a new operator $K'$ acting in ${\cal S '}$.  Let
\beq
b = K a.
\label{27}
\eeq
Notice that
\beq
<b |\phi_i > = - \frac{C_i}{g_i} <a|a> = - \frac{C_i}{g_i} .
\label{28}
\eeq
We try
$$
K' = K - |b><a| - |a><b| + C |a><a|
$$
where $C$ will be chosen so that
\beq
K' a = 0.
\label{29}
\eeq
We have
$$
K'|a> = |b> - |b> - <b|a> |a> + C | a>,
$$
and hence we take
\beq
C = <b|a> = <a |K| a>
\label{30}
\eeq
$K'$ is Hermitian like $K$, and we get
\bea
g_i K' |\phi_i > & = & g_i |K|\phi_i > - g_i <b|\phi_i > |a> \nonumber \\
& = & g_i |K|\phi_i > + C_i |a> .\nonumber
\eea
Hence
\beq
g_i K' |\phi_i > = |\phi_i >
\label{31}
\eeq
which is \underline{homogeneous}.

To get a bound on the number of bound states of type II, we have to get a
bound on trace
$K'^2$ (not surprisingly, trace $K'$ is divergent).  It is a lengthy but
straightforward
exercise to calculate that trace, which gives
\beq
N_{II} < \, {\rm tr} \, K^2 - 2 <a|K^2 |a> + (<a|K|a>)^2
\label{32}
\eeq
The last two terms give an overall negative contribution.  The first term
is the same as
the one appearing in $N_I$.  It is easy to see that the right-hand side of
(32) is
independent of the scale parameter $k_0$ entering into the kernel $K$.
Finally, let us
notice that the treatment of case II \underline{contains} case I because,
in the argument,
it has never been said that $C_i \not= 0$.  Equation (31) holds
irrespective of whether
$C_i = 0$ or $C_i \not= 0$.  Notice that the bound on $N_{I}$ is larger
than the bound on
$N_{II}$. Therefore the bound on $N_I$ becomes completely obsolete.

\section{Concluding Remarks}

If we include the bound state with evanescent energy for zero coupling
constant, we get
the bound
$$
N < 1 + \, {\rm tr} \, K^2 - 2 <a |K^2|a> + (<a|K|a>)^2.
$$
Dropping the last two terms still gives a scale-dependent bound - which can
be minimized with respect to
the scale - which precisely appears in condition (2), itself following from
the linear condition (3) as
shown in the Appendix.

Conditions (2) and (3) both allow a potential behaving like
$$
\frac{1}{r^2 (\ell n r)^{3+\epsilon}}
$$
at infinity, with local singularities not worse than
$$
- \frac{1}{|r - r_0|^2 (| \ell n |r - r_0||)^{2+\epsilon}},
$$
$\epsilon$ positive, arbitrarily small.  Both conditions are violated for
$\epsilon < 0$. However, we shall see in the
Appendix that (2) is definitely weaker than (3).

Our bound has the merit of being valid for the non-central case, which, as
we said in the Introduction, is
important for solid-state physics.  However, for a potential $gV$, it
behaves like $g^2$ for large $g$,
while in Ref. \cite{aaa}, in the central case, we get a bound behaving like
$g$.  In Ref. \cite{aaa}
we make a conjecture which is very far from being proved, but clever
mathematical physicists might prove it
or something similar.  The present work should be considered only as a
first step which could possibly give
reasonable results for not too large $g$.

\noindent
{\bf Acknowledgements}

We are grateful to P. Sabatier for suggesting the use of the non-linear
expression defined by (2) for potentials in two
dimensions.  Our work was stimulated by discussions with our colleagues K.
Chadan and N.N. Khuri.  One of us (T.T.W.)
would also like to thank the CERN Theoretical Physics Division for
hospitality.  This paper was put in final form while
one of us (A.M.) was visiting the ``Institut des Hautes Etudes
Scientifiques", whose hospitality is acknowledged.

\setcounter{section}{0}
\setcounter{equation}{0}
\renewcommand{\thesection}{Appendix:}
\renewcommand{\theequation}{A. \arabic{equation}}

\section{\large{ \bf Comparison of condition (2)  and condition (3)} }

Condition (2) is
\beq
I = \int d^2 x d^2 y \, V^-(x) (\ell n |x-y|)^2 \, V^- (y) < \infty
\label{A1}
\eeq
Condition (3) is a set of two conditions:
\beq
\int d^2 x \, \left ( \ell n (2+|x|) \right )^2 \, V^- (x) < \infty
\label{A2}
\eeq
\beq
\int d^2 x \, V_R (|x|) \, \ell n^-(|x|) < \infty .
\label{A3}
\eeq
In (A.3) we use:
\bea
- \ell n^- (|x|) & = & 0 \; {\rm if} \;|x| > 1 \nonumber \\
& = & - \ell n |x| \; {\rm if} \;|x| < 1 . \nonumber
\eea

$- V_R (|x|)$, the circular decreasing rearrangement of $V^- (x)$.  Since
this notion is not very well known
among physicists, let us remind the reader that $V_R (|x|)$ is a decreasing
function of $|x|$, such that
$$
\mu (V_R (|x|) > A) = \mu (V^- (x) > A), \, \forall A,
$$
where $\mu$ is the Lebesque measure.  In more familiar terms, the
rearranged Mont Blanc would be a mountain
with axial symmetry, with a single peak, such that the surface  between the
level lines would be the same as
the surface between the level lines of the original Mont Blanc (rather
awfully dull!).

We shall prove first that the convergence of $I$ in (A.1) follows from the
convergence of (A.2) and (A.3).
More exactly, we shall get an explicit bound on (A.1) in terms of (A.2) and
(A.3).
We write
\beq
I = I_+ + I_-
\label{A4}
\eeq
with
\beq
I_+ = \int d^2x d^2y \, V^- (x) \, (\ell n^+ |x-y|)^2 \, V^- (y)
\label{A5}
\eeq
\beq
I_- = \int d^2x d^2y \, V^- (x) \, (\ell n^- |x-y|)^2 \,V^- (y)
\label{A6}
\eeq
$\ell n^-$ has already been defined.  $\ell n^+ (a) = \ell  n  a$ for $a
\geq 1$, $= 0$ for $a < 1$. It is
elementary to get a bound on $I_+$ from (A.2) only.  Indeed,
$$
0 < \ell n^+ |x-y| < \ell n^+ (|x|+|y|) < \ell n (2 +|x|) + \ell n (2 + |y|)
$$
and thus
\beq
\left ( \ell n^+ |x-y| \right)^2 < 2 \left [ (\ell n 2 + |x|)^2 + (\ell n
(2+|y|))^2 \right ]
\label{A7}
\eeq
Hence
\beq
I_+ < 4 \int d^2 x \, V^- (x) \; \int d^2 y \, V^- (y) \, (\ell n (2+|y|))^2 .
\label{A8}
\eeq
The convergence of the right-hand side of (A.8) follows directly from (A.2).

Concerning $I_-$, we use a rearrangement inequality due to Luttinger and
Friedberg \cite{jj}, which says
\beq
\int \int \, A (x) \, B (|x-y|) \, C (y) d^2x d^2y \leq \int \, A_R (|x|)
\, B_R (|x-y|) \, C_R (|y|) \,
d^2x d^2y
\label{A9}
\eeq
where $A, B, C$ are non-negative functions and $A_R, B_R, C_R$ are their
decreasing rearrangements.  Since
$\ell n^-$ and $(\ell n^-)^2$ are decreasing functions of their argument
they are their own rearrangement.
Hence
\beq
I_- < \int \, d^2x d^2 y \; V_R (|x|) \left ( \ell n^- (|x-y|) \right )^2
\, V_R (|y|)
\label{A10}
\eeq

In (A.10), we can carry out first the angular integration, the angle
($\vec{x}, \vec{y})$ appearing only in
$\ell n^-$.  However, to be able to do that easily we have to sacrifice
some information, i.e., use
$(\ell n^- |x-y|)^2 \leq (\ell n (|x-y|))^2$.  We have to calculate
$$
\int \, \frac{d \theta}{2\pi} \, \left ( \ell n (|x|^2 + |y|^2 -
2|x||y| \, \cos \theta) \right )^2 .
$$
We have
$$
\ell n (|x|^2 + |y|^2 - 2 |x||y| \, \cos \theta ) =
\ell n (|x| - |y| e^{i \theta}) + \ell n (|x| - |y| e^{-i \theta}).
$$
Assume $|x| > |y|$.  Then we get
\beq
\ell n (|x|^2 + |y|^2 - 2 |x||y| \, \cos \theta ) =
2 \left [ \ell n |x| - \Sigma \left( \frac{|y|}{|x|} \right )^n \;
\frac{\cos n \theta}{n} \right ]
\label{A11}
\eeq
Hence, if $|x| > |y|$, using the orthogonality of the $\cos \, n \, \theta$:
$$
\int \, \frac{d \theta}{2\pi} \, (\ell n (|x|^2 + |y|^2 - 2 |x||y| \, \cos
\theta ))^2 =
4 ( \ell n |x|)^2 + 2 \sum^{\infty}_{n=1} \left( \frac{|y|}{|x|} \right
)^{2n} \,
\frac{1}{n^2}
$$
We see a dilogarithm, or Spence function, appearing on the right-hand side.
However, we only need to
notice that
\beq
{\rm sup}_{|y| \leq |x|} \; \int \frac{d\theta}{2\pi} \,
\left ( \ell n |x|^2 + |y|^2 - 2 |y||x| \cos \theta \right )^2
= 4 (\ell n |x|)^2 + 2 \sum^{\infty}_{n=1} \, \frac{1}{n^2} =
4 (\ell n |x| )^2 + \frac{\pi^2}{3} .
\label{A12}
\eeq
In this way we get
$$
I_{-} < (2\pi)^2  \times 2 \int_{|x|>|y|} |x|d|x| \, |y|d|y| \, V_R (|x|)
V_R (|y|) \;
\left [ 4 (\ell n |x|)^2 + \frac{\pi^2}{3} \right ]
$$
Again, we split the integral into
\bea
& 32 \pi^2 &\, \int_{|x|>|y|} \, |x|d|x| \, |y|d|y| \, V_R(|x|) V_R(|y|)
(\ell n^- |x|)^2 \nonumber \\
+ & 32 \pi^2 &\, \int_{|x|>|y|} \, |x|d|x| \, |y|d|y| \, V_R(|x|) V_R(|y|)
(\ell n^+ |x|)^2 \nonumber \\
+ & \frac{4\pi^4}{3} &\left [ \int |x|d|x| \, V_R (|x|) \right ]^2
\label{A13}
\eea
In the first term of (A.13) we can replace $(\ell n^- (|x|))^2$ by $\ell
n^-|x| \, \ell n^- |y|$, since
$|x| > |y|$ and since $\ell n^-$ is decreasing.

In the second term, we can drop the restriction $|x| > |y|$ and notice that
$$
\int d^2x V_R (|x|) \, \left ( \ell n^+ (|x|) \right )^2 < \int d^2x V_-(x)
\left ( \ell n^+ (|x|) \right
)^2 .$$
Indeed,
$\int d^2x A_R (x) \phi (|x|)$, where $\phi (|x|)$ is
\underline{increasing}, is less than $\int d^2x \,
A(x) \, \phi(|x|)$.  Suppose that $\phi (|x|) \to L$.  Then
$$
\int d^2x A_R (x) \phi (|x|) = \int d^2x A_R (x)L - \int d^2x A_R (x)
(L-\phi(|x|)).
$$
$L - \phi(|x|)$ is its own decreasing rearrangement and following the
well-known properties
$$
\int A_R B_R d^2x \geq \int A(x) B(x) d^2x
$$
 and
$$
\int A_R (|x|) d^2x = \int A(x) d^2x,
$$
we get the desired
property.  If $L$ is infinite, we can use a limiting procedure.  Finally,
we get
\bea
I_- & < & 16 \pi^2 \left [ \int d^2 x \, V_R (|x|) \, \ell n^- |x| \right
]^2 \nonumber \\
& + & 32 \pi^2  \int d^2 x \, V_-(x) \, \int d^2 y \, V_-(y) (\ell n
(2+|y|))^2 \nonumber \\
& + & \frac{4 \pi^4}{3} \left [ \int d^2 x \, V_-(x) \right ]^2
\label{A14}
\eea
From (A.2) and (A.3) we see that $I_-$ is bounded.  This concludes the proof.

One question is: can we go in the opposite direction? Assume that we know
that (A.1) holds.  There exists
certainly a region $|x -x_o |< R$ where ${\rm Inf} \, V^- = m > 0$.  If
such a region did not exist, $V^-$
would be zero almost everywhere!  So
\beq
I > \pi R^2 m \;
\int_{|x|>|x_o|+R+2} V^- (x) \, \ell n [2+|x|]^2 d^2 x
\label{A15}
\eeq
Now we choose $y_o > 4 + |x_o| + R + 2$ such that
$$
{\rm Inf}_{|y-y_o| < R'} \, V_-(y) = m' > 0 ,
$$
 then
\beq
I > \pi R'^2 m' \;
\int_{|x|<|x_o|+R+2} \, V^- (x) (\ell n (4-R'))^2 d^2 x.
\label{A16}
\eeq
This proves that the convergence of (A.1) implies the convergence of (A.2).

It is not possible to deduce (A.3) from (A.1) because (A.3) involves $V_R$
and (A.1) does not. However, in
practice the conditions are very similar.  Nevertheless, the following
example shows that (A.3) is stronger
than (A.1), even for a potential which does not need rearrangement: take
the central potential
\bea
V(|x|) & = & - \frac{1}{|x|^2 |\ell n x|^2 (\ell n |\ell n |x||)^{\gamma}}
\; {\rm for} \; |x| <
\frac{1}{2e}, \nonumber \\
& = & 0 \; {\rm for} \; |x| \geq \frac{1}{2e}.
\label{A17}
\eea
For $\gamma \leq \frac{1}{2}$ (A.1) and (A.3) are divergent,
 for $\frac{1}{2} < \gamma \leq 1$ (A.1) is
convergent and (A.3) is divergent,
 for $\gamma > 1$ (A.1) and (A.3) are convergent.

\end{document}